\documentclass[12pt]{article}
\usepackage{psfig,epsf}
\usepackage{a4wide,graphicx}
\usepackage{cite}

\begin{document}

\title{Chiral Dynamics of scalar mesons: radiative $\phi$ decay and $\sigma$ in the
medium through $\pi^0 \pi^0$ nuclear photoproduction}

\maketitle

\begin{center}

\author{
E.~Oset, L.~Roca, M.~J.~Vicente Vacas and J.~Palomar\\
{\small Departamento de F\'{\i}sica Te\'orica and IFIC,
Centro Mixto Universidad de Valencia-CSIC,} \\ 
{\small Institutos de
Investigaci\'on de Paterna, Aptdo. 22085, 46071 Valencia, Spain}\\ 
}

\end{center}

\begin{abstract}
In order to assess the relevance of chiral dynamics in the scalar sector we
address two recent problems: radiative decay of the $\phi$, for which there
are quite recent data from Frascati, and the modification of the $\sigma$
properties in the nuclear medium seen through the $\pi^0 \pi^0$
photoproduction in nuclei.
\end{abstract}


\section{Introduction}

The radiative decays of the $\phi$ into $\pi^0 \pi^0 \gamma$ and 
$\pi^0 \eta \gamma$ have been the subject of intense study
\cite{greco,franzini,colangelo,achasov,bramon5,lucio,uge}. 
One of the main reasons for this is the hope that one can get much information
about the nature of the $f_0(980)$ and $a_0(980)$  resonances .  The nature of the scalar meson 
resonances has generated a large debate  \cite{kyoto}, with new ideas brought by the 
claim that  these resonances are dynamically generated from multiple
scattering with the ordinary chiral
Lagrangians  \cite{npa,iam,nsd}.

   These two reactions involving the decay of the $\phi$ are 
special. Indeed, the $\phi$ does not decay into two pions
because of isospin symmetry. 
  But we can bypass this by allowing the $\phi$ to decay into
two charged kaons (with a photon attached to one of them) and
the two kaons scatter giving rise to the two pions (or
$\pi^0 \eta$).  The loop which appears diagrammatically is
proved to be finite using  arguments of gauge invariance 
\cite{pertiou,close,bramon5}.  The radiative $\phi$ decay
through this mechanism was studied in \cite{bramon5} and the
results of lowest order chiral perturbation theory ($\chi
PT$) were used to account for the $K^+ K^- \to \pi^0 \pi^0$
transition.  Since the  chiral perturbation theory $K^+ K^-
\to \pi^0 \pi^0$ amplitude does not account for the
$f_0(980)$, the excitation of this resonance has to be taken
in addition, something that has been done more recently using
a linear Sigma-model in \cite{bramonnew}.  

  The work of \cite{uge} leads to the excitation of the 
$f_0(980)$  in the  $\pi^0 \pi^0$ production, or the
$a_0(980)$ in $\pi^0 \eta$ production in a natural way, since
the use of unitarized chiral perturbation theory ($U \chi
PT$), as in \cite{npa}, generates automatically those
resonances in the meson meson scattering amplitudes and one
does not have to introduce them by hand. 

The experimental situation has also experienced an impressive
progress recently.  To the already statistically significant 
experiments at Novosibirsk \cite{novo1,novo2,novo3} one has
added the new, statistically richer,
experiments at Frascati
\cite{frascati1,frascati2} which allow one to test models
beyond just the qualitative level.   In this sense although
the predictions of the work of \cite{uge}, using $U \chi PT$
with no free parameters, provided a good agreement with the
experimental data of \cite{novo1,novo2,novo3}, thus settling
the dominant mechanism as that coming from chiral kaon loops
from the $\phi \to K^+ K^-$ decay, the new and more  precise
data leave room for finer details which we evaluate in this
paper.

In addition to the mechanisms discussed before we have sequential $V \to VP \to PP \gamma$ process. This mechanism
is known to provide the $\omega \to \pi^0 \pi^0 \gamma$
radiative width with accuracy \cite{grau} and has been
further extended to study  $\rho \to \pi^0 \pi^0 \gamma$ and
other radiative decays in \cite{escribano,palomar}.

  Another novelty of the present work is the  consideration
of sequential mechanisms involving the exchange of an
intermediate axial-vector meson ($J^{PC}=1^{++}$ or
$1^{+-}$), both producing directly the final meson pair or
through the intermediate production of kaons which undergo
collisions and produce these mesons.

   All the mechanisms considered here contribute moderately,
but appreciably, to the $\phi$ radiative width. The inclusion
of all these mechanisms leads to results compatible with the
experimental data of Frascati, particularly if theoretical
uncertainties are considered, which is something also done in
the present work.

   The good agreement with experiment is reached in spite of
having in our approach a width for the $f_0(980)$ very small,
of the order of 30 MeV, in apparent contradiction with the
"visual"  $f_0(980)$ width in the experiment, which looks
much larger. The reason for this has been recently discussed
in  \cite{achasovq,pennington} and stems from the fact that,
due to gauge invariance, the amplitude for the process
contains as a factor the momentum of the photon, which grows
fast as we move down to smaller invariant masses from the
mass of the  $f_0(980)$ where the photon momentum is very
small.  This distorts the shape of the resonance, making it
appear wider.  Our approach, which respects
gauge invariance, introduces automatically this factor in the
amplitudes.

   We shall see that there is some discrepancy of the
theoretical results  with the data at small invariant
masses.  We shall discuss this feature, realizing that the
results resemble very much the raw data, before the analysis
is done to subtract the contribution of $\omega \pi^0$ and to
correct for the experimental acceptance. Furthermore, some of the
assumptions made in the analysis of  \cite{frascati1} might be 
questionable.

\section{The $\phi\to\pi^0\pi^0\gamma$ decay}

\subsection{Kaon loops from $\phi \to K^+ K^-$ decay
\label{sec:chiral_loops}}

\begin{figure}
  \includegraphics[width=.5\textwidth]{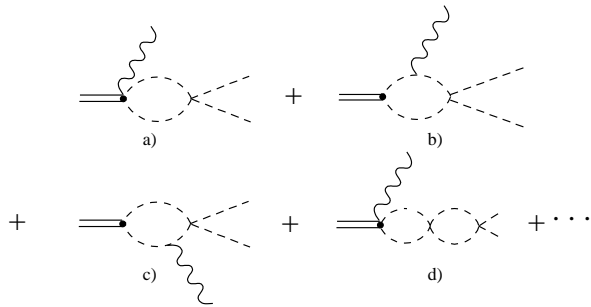}
  \caption{Loop diagrams included in the chiral loop contributions. The
intermediate states in the loops are $K^+K^-$.}
\label{looplot}
\end{figure}

  The mechanism for radiative decay using the tensor formulation for the vector
mesons have been discussed in \cite{neufeld,uge} and we briefly summarize it
here. The diagrams considered are depicted in Fig.~\ref{looplot}, where the 
loops contain 
$K^+ K^ -$. The vertices needed for the diagrams are obtained from
the chiral Lagrangian for vector meson resonances of ref. \cite{ecker}, assuming ideal mixing between the $\phi$ and $\omega$ mesons.

Using arguments of gauge invariance it was proved in \cite{pertiou,close} that the loop
functions are convergent and in \cite{oller} that the meson meson amplitude
factorizes outside the loop integral with on shell value. This
information is of much value, since it allows to factorize the
on shell meson meson amplitude outside the loop integral. The
amplitude for the process is given by 

\begin{equation} \label{eq:loopsuge}
t=-\sqrt{2}\frac{e}{f^2}\epsilon(\phi)\cdot \epsilon(\gamma) \left[
M_{\phi}G_V\tilde G(M_I)+q%
\left(\frac{F_V}{2}-G_V\right)G(M_I) \right]
t_{K^+K^-,\pi^0\pi^0}
\end{equation}
where $f=92.4\textrm{ MeV}$ and 
$\tilde{G}$ is the convergent loop function of
\cite{pertiou,close}. On the other hand,
$G(M_I)$ is the ordinary loop function of two meson
propagators which appears in the study of the meson meson
interaction in \cite{npa} and which is regularized there with
a cut off of the order of $1\,$GeV. In 
Eq.~(\ref{eq:loopsuge}) the  $t_{K^+K^-,\pi^0\pi^0}=
\frac{1}{\sqrt{3}}t^{I=0}_{K\overline{K},\pi\pi}$ is the
transition amplitude with the iterated loops
implicit in the coupled channels Bethe Salpeter equation (BS)
 obtained  in
\cite{npa}. 
The parameters $F_V$, $G_V$, for the vector mesons are
obtained from their decay into $e^+e^-$, $\mu^+\mu^-$ or two
mesons. 
 We take for the calculations 
 $F_V=156\pm 5\textrm{ MeV}$ and  $G_V=55\pm 5\textrm{ MeV}$ (see \cite{ourphirad} for a discussion on these vales).

\subsection{Sequential vector meson exchange mechanisms
\label{sec:VMDtree}}

Following the lines of \cite{escribano,palomar} in the
study of $\rho$ and $\omega$ radiative decays and the more
recent of \cite{achasovlast,lucio} in the $\phi$ decay,
we also include these mechanisms here. They are depicted in
Fig.~2, where we explicitly assume that the
$\phi\to\rho^0\pi^0$ proceeds via the $\phi-\omega$ mixing.

\begin{figure}
  \includegraphics[width=12cm]{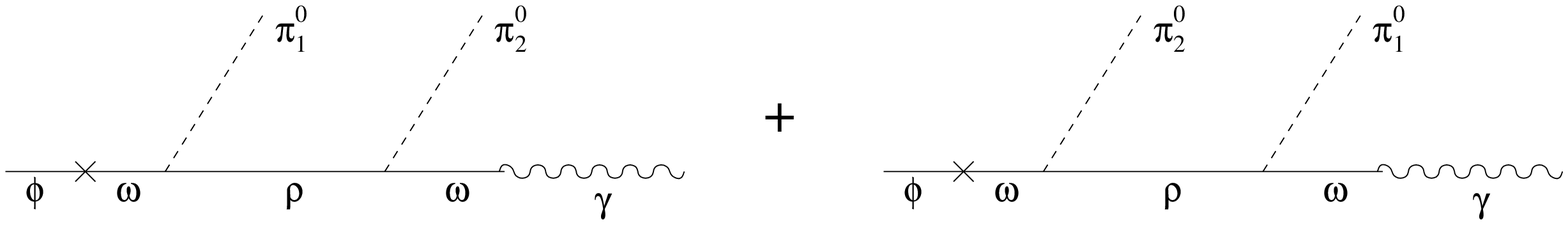}
  \caption{\rm 
  Diagrams for the tree level VMD mechanism.}
 \label{fig:VMD_tree}
\end{figure}

In order to evaluate these diagrams
 we use the same Lagrangians as in \cite{bramon5,escribano}.
In addition we must use the Lagrangians producing the 
$\phi$-$\omega$ mixing. We use the formalism of \cite{urech}

\begin{equation}  
{\cal L}_{\phi\omega} = \Theta_{\phi\omega}
\,\phi_{\mu}\omega^{\mu}
\end{equation}
\noindent 
which means that the diagrams of Fig.~\ref{fig:VMD_tree}
 can be evaluated
assuming the decay of the $\omega$ (with mass $M_{\phi}$)
multiplying the amplitude
 by $\tilde{\epsilon}$ (the measure of the
$\phi$-$\omega$ mixing) given by $\tilde{\epsilon} = 
\Theta_{\phi\omega}/(M_{\phi}^2-M_{\omega}^2)$

Values of $\Theta_{\phi\omega}$ of the order of
$20000-29000\textrm{ MeV}^2$ are quoted in
\cite{kozhe} which are compatible with 
$\tilde{\epsilon}=0.059\pm 0.004$
used in \cite{bramonepsi}
\footnote{Note that in \cite{bramonotro} a different sign for
$\tilde{\epsilon}$ is claimed. This is
 actually a misprint and the
results of that paper are calculated with
 $\tilde{\epsilon}>0$ \cite{bramonprivate}.}
 which is the value used here.

The amplitude for the
$\phi(q^*)\to\pi^0_1(p_1)\pi^0_2(p_2)\gamma(q)$ decay
corresponding to the diagrams of Fig.~2 is given
by

\begin{equation}
t=-{\cal C}\tilde{\epsilon}
\frac{2\sqrt{2}}{3}\frac{egf^2G^2}{M_{\omega}^2}
\left[\frac{P^2\{a\}+\{b(P)\}}{M_{\rho}^2-P^2-iM_\rho
\Gamma_\rho(P^2)}
+\frac{P'^2\{a\}+\{b(P')\}}{M_{\rho}^2-P'^2-iM_\rho
\Gamma_\rho(P'^2)}
\right]
\label{eq:VMDtree1}
\end{equation}
where $P=p_2+q$, $P'=p_1+q$ and $\{a\}$, $\{b(P)\}$ given in \cite{escribano}. 

At this point it is worth mentioning that the theoretical
expression for the $V\to P\gamma$
decay widths $\Gamma_{V \to P\gamma}=\frac{4}{3} \alpha
 C_i^2 \left(\frac{G g f^2}{M_\rho M_V}\right)^2k^3$, 
with $C_i$ $SU(3)$ coefficients given 
in Table I of \cite{roca}, gives slightly different results
to the experimental values from \cite{pdg}.
 For this reason we can follow a similar procedure to that 
used for the $\eta\to\pi^0\gamma\gamma$ decay in \cite{roca}
where the $C_i$ coefficients were normalized
so that the theoretical $V\to P\gamma$ decay widths agree with
experiment. In the $\phi\to\pi^0\pi^0\gamma$ reaction
this procedure results in including in   
Eq.~(\ref{eq:VMDtree1}) a normalizing
factor ${\cal C}=0.869\pm 0.014$, obtained considering the
$V\to P\gamma$ reactions shown in Table I of \cite{roca}.

\subsection{Pion final 
state interaction in the sequential vector
meson mechanism \label{sec:VMDpiloops}}

Since the $\pi\pi$ interaction is strong in the region of
invariant masses relevant in the present reaction we next
consider the final state interaction of the pions in the
sequential vector meson mechanism. 

\begin{figure}
  \includegraphics[width=15cm]{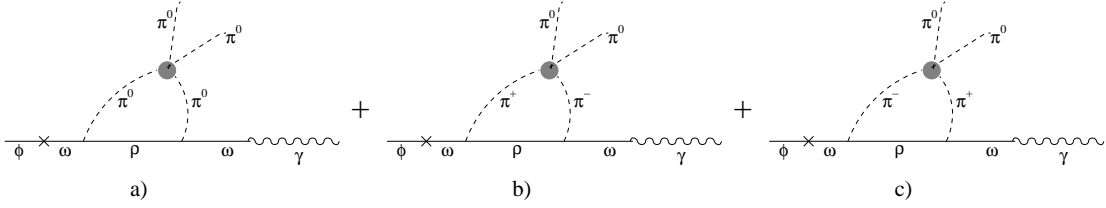}
\caption{\rm 
VMD diagrams with final state interaction of pions}
\label{fig:VMD_loop_pions}
\end{figure}

 We must take into account the loop function of
Fig.~\ref{fig:VMD_loop_pions}a,
 but on the same footing we must also consider
those of Fig.~\ref{fig:VMD_loop_pions}b
and \ref{fig:VMD_loop_pions}c, where charged pions are
produced and allowed to interact to produce the $\pi^0\pi^0$
final state. The thick dot in Fig.~\ref{fig:VMD_loop_pions}
means that one is considering the full $\pi\pi\to\pi\pi$
t-matrix, involving the loop resummation of the BS equation of
ref.~\cite{npa} and not just the lowest order 
$\pi\pi\to\pi\pi$ amplitude.

In order to evaluate those diagrams we must calculate the
loop function with a $\rho$ and two pion propagators. First
let us note that due to isospin symmetry the
$\omega\rho^0\pi^0$ coupling is the same as the
$\omega\rho^+\pi^-$ or  $\omega\rho^-\pi^+$. Next, given the
structure of the terms in Eqs.~(\ref{eq:VMDtree1})  we must evaluate the loop integrals 

\begin{equation}   \label{eq:loop3mesons}
i\int\frac{d^4P}{(2\pi)^4} \,P^{\mu}P^{\nu}
\frac{1}{P^2-M_V^2+i\epsilon}
\,\frac{1}{(q^*-P)^2-m_1^2+i\epsilon}
\,\frac{1}{(q-P)^2-m_2^2+i\epsilon}
\end{equation}

for which we evaluate first the $P^0$ integral analytically and then the
three momentum integral numerically with a cut off of 1 GeV.

\subsection{Kaon loops in the sequential vector
 meson mechanisms \label{sec:VMDKloops}}

Next we consider the diagrams analogous to those in Fig.~3 but
with kaons and $K^*$ in the intermediate states.

Note that the $\phi VP$ vertices
are now not OZI forbidden. They come from the Lagrangian of 
refs. \cite{bramon5,escribano}, and all the four $\phi K K^*$ vertices have the same
strength.

\subsection{Sequential axial vector meson mechanisms}

Since the mass of the $\phi$ is around $250\textrm{ MeV}$
higher than the $\rho$ mass, and we are considering sequential
vector meson mechanisms with $\rho$ or $K^*$ exchange, we
should pay attention to the analogous mechanisms involving
vector mesons with a similar mass difference with the $\phi$
on the upper side and these are the axial and vector mesons
with $J^{PC}=1^{+-}$ or $1^{++}$ (see
Table~1). Therefore,  the $b_1$ or $a_1$ axial vector
 mesons and the
$K_{1B}$, $K_{1A}$ strange axial vector mesons will play the
role of the $\rho$ or the $K^*$ in former diagrams.

\begin{table}[htbp]
\begin{center}
\begin{tabular}{|c||c||c|c|}\hline 
$J^{PC}$ &$I=1$  &$I=0$ & $I=1/2$ \\ \hline \hline  
 $1^{+-}$  & $b_1(1235)$  & $h_1(1170)$, $h_1(1380)$
    &$K_{1B}$ \\ \hline  
 $1^{++}$  & $a_1(1260)$  & $f_1(1285)$, $f_1(1420)$
    &$K_{1A}$ \\ \hline     
\end{tabular}
\end{center}
\caption{Octets of axial-vector mesons.}
\end{table}

Because of the $C$
parity of the states, the Lagrangians for
the axial-vector--vector--pseudoscalar couplings have
the structure of $<B\{V,P\}>$ for the $b_1$ octet and 
$<A[V,P]>$ for the octet of the $a_1$ \cite{gatto}, where
the $<>$ means $SU(3)$ trace. In the last expressions $V$ and
$P$ are the usual vector and pseudoscalar $SU(3)$ matrices
respectively and $B$ and $A$ are axial vector $SU(3)$ matrices given in \cite{ourphirad}.

 In addition one has to consider 
an approximate $50\,$\%  mixture of the
$K_{1B}$ and $K_{1A}$ states to give the physical $K_1(1270)$
and $K_1(1400)$  states \cite{gatto,suzuki,carnegie}.

We have modified the
original Lagrangian of \cite{gatto} to treat the
vector fields in the tensor formalism of \cite{ecker}. This
formalism has the advantage that without basically changing
the rates of the $A\to VP$ decays, one deduces the coupling of
the $a_1$ to $\pi\gamma$ using vector meson dominance through
$a_1\to\pi\rho\to\pi\gamma$, with an amplitude which is gauge
invariant  and which is in agreement with the chiral structure
of \cite{ecker} for  the $a_1\to P\gamma$ couplings and with
the experiment. Details are given elsewhere in \cite{axial}.

We hence use the Lagrangians \cite{axial}

\begin{eqnarray}  \label{eq:LBLA}
{\cal L}_{BVP}&=&\tilde{D} <B_{\mu\nu}\{V^{\mu\nu},P\}> \\
\nonumber
{\cal L}_{AVP}&=&i\tilde{F} <A_{\mu\nu}[V^{\mu\nu},P]> 
\end{eqnarray} 
where the $i$ factor in front of the $\tilde{F}$ is needed
in order ${\cal L}_{AVP}$ to be hermitian.

In Eq.~(\ref{eq:LBLA}) the fields 
$W_{\mu\nu}\equiv A_{\mu\nu}$, $B_{\mu\nu}$ are normalized
such that 
\begin{equation}
<0|W_{\mu\nu}|W;P,\epsilon>=
\frac{i}{M_W}\left[ P_\mu\,\epsilon_\nu(W)-P_\nu\, 
\epsilon_\mu(W)\right]
\end{equation}

The physical $K_1(1270)$ and $K_1(1400)$, with a mixture
around $45$ degrees
\footnote{It is worth mentioning that in \cite{suzuki,axial} two
more possible solutions for the mixing angle around $30$ and
$60$ degrees were found. This uncertainty will be taken into
account in the evaluation of the error band in our final
results.}
 found in
\cite{gatto,suzuki,carnegie,axial}, can be expressed, in
terms of the $I=1/2$ members of the $1^{+-}(1^{++})$ octets, 
$K_{1B}(K_{1A})$, as

\begin{eqnarray}
\nonumber
K_1(1270)=\frac{1}{\sqrt{2}}(K_{1B}-iK_{1A}) \\
K_1(1400)=\frac{1}{\sqrt{2}}(K_{1B}+iK_{1A})
\end{eqnarray} 

With the values for  $\tilde{D}=-1000\pm 120$  MeV and 
$\tilde{F}=1550\pm 150$ MeV, very similar to those found in
\cite{gatto,suzuki,carnegie}, we are able to describe all the
$A\to VP$ decays plus the radiative decays of the
$a_1\to\pi\gamma$ \cite{axial}.

Once again the $\phi$ sequential decay
at tree level through $b_1$ exchange
is OZI violating and  found to be negligible.

\subsection{Kaon loops from sequential axial vector meson mechanisms}

The relevant mechanisms involving axial-vectors
are those  in which
 $K$, $\bar{K}$ are created and
through scattering lead to the final $\pi^{0}\pi^{0}$ state, having also one of the $K_1$ resonances as intermediate state. These are not OZI
forbidden and have a nonnegligible contribution.

Since we are using the tensor formulation for the vector
mesons this forces us to use the tensor coupling of the photon
to the vector mesons obtained from the $F_V$ term of
the chiral Lagrangian of ref. \cite{ecker}:

\begin{equation}
{\cal L}_{V\gamma}=-e\frac{F_V}{2}\lambda_V V_{\mu\nu}^0
(\partial^\mu A^\nu - \partial^\nu A^\mu)
\end{equation}

with $\lambda_V=1,\frac{1}{3},-\frac{\sqrt{2}}{3}$ for 
$V=\rho,\omega,\phi$ respectively.

\section{Results for the $\phi\rightarrow\pi^{0} \pi^{0} \gamma$ decay}

\begin{figure}
  \includegraphics[width=7cm,angle=-90]{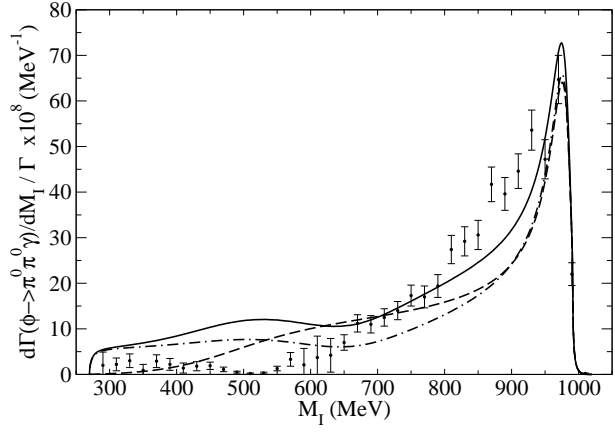}
\caption{\rm 
Different contributions to the two pion invariant mass
distributions of the $\phi\to\pi^0\pi^0\gamma$ decay:
Dashed line: chiral loops of Fig.~\ref{looplot}.
Dashed-dotted line: chiral loops of Fig.~\ref{looplot} + 
sequential VMD and its final
state interaction.
Solid line: idem plus the contribution of the mechanisms
involving axial-vector mesons, (full model).
}
\label{fig:res6}
\end{figure}

In Fig. \ref{fig:res6} we show the results of the different
contributions. We shoud say that the loops of the sequential vector meson
mechanisms involving kaons are relatively important but there is a strong
cancellation between the mechanisms with a $\phi$ and an $\omega$ attached to
the photon.

\begin{figure}
  \includegraphics[width=6cm,angle=-90]{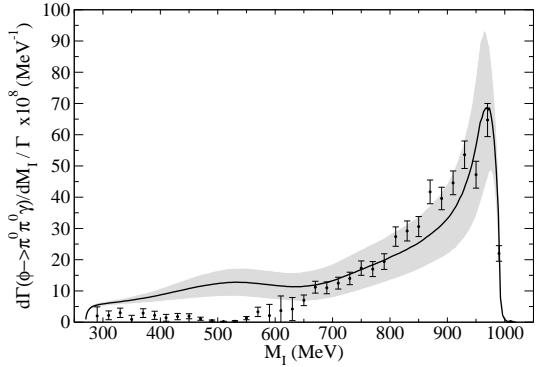}
\caption{\rm 
Final results for the $\pi^0\pi^0$ invariant mass distribution
for the $\phi\to\pi^0\pi^0\gamma$ decay with the theoretical
error band. Experimental data from \cite{frascati1}.
}
\label{fig:res7}
\end{figure}

Up to now, all the curves shown in the figures have been
calculated using the central values of the parameters without
considering the uncertainties in their values. 
In Fig.~\ref{fig:res7} we
show the final result but including an evaluation of the error
band due to the uncertainties in the parameters of the model.
This error band has been calculated implementing a Monte
Carlo gaussian sampling of the parameters within their
experimental errors. The parameters of the model which
uncertainties are relevant in the error analysis
are shown in Table \ref{tabla2}.

\begin{table}[tbp]
\begin{center}
\begin{tabular}{|c|c|c|c|}\hline 
${\cal C}$ &$\tilde{\epsilon}$  &$G_V$ (MeV) &
 $F_V$ (MeV) \\  
 $0.869\pm 0.014$  & $0.059\pm 0.004$  & $55\pm 5$&$156\pm 5$
   \\ \hline & & &  \\[-0.4cm]
 $f_\pi$  (MeV)& $\Lambda$ (MeV) & $\tilde{D}$ (MeV)&
  $\tilde{F}$ (MeV)
   \\
  $92.4\pm 3$\% & $1000\pm 50$ & $-1000\pm 120$&$1550\pm150$
  \\ \hline     
\end{tabular}
\end{center}
\caption{Parameters which uncertainties are relevant in
 the error analysis. The $f_\pi$  and $\Lambda$ are the $f_\pi$
constant and cutoff of the momentum integral respectively 
in the loops involved  in the unitarized
 meson-meson rescattering.}
 \label{tabla2}
\end{table}

The errors in $f_\pi$ and $\Lambda$ assumed
 in the calculations
have been chosen such that the quality of the fit to the
$\pi\pi$ phase shifts along the lines of \cite{npa} is still
acceptable within experimental errors.

The parameter with the larger contribution to the error band
turns out to be the $G_V$ since the largest contribution,
chiral kaon loops form $\phi\to K^+K^-$ decay, is roughly
proportional to $G_V$ (up to the term with
$q(\frac{F_V}{2}-G_V)$ in Eq.~(\ref{eq:loopsuge}) which would
be zero within some vector meson dominance hypotheses
\cite{ecker} and is small with our set of parameters).

The total width and branching ratio obtained in the present
work are

\begin{eqnarray}
BR(\phi\to\pi^0\pi^0\gamma)&=&(1.2\pm 0.3)\times 10^{-4}
\end{eqnarray}
to be compared with the experimental values 

$BR^{exp}(\phi\to\pi^0\pi^0\gamma)
=(1.22\pm 0.10\pm 0.06)\times 10^{-4}$  \cite{novo1}, 
$(0.92\pm 0.08\pm 0.06)\times 10^{-4}$,\cite{novo2}, 
$(1.09\pm 0.03\pm 0.05)\times 10^{-4}
$ \cite{frascati1}.

In Fig.~\ref{fig:res7} we can see that our results, considering the error
band, fairly agrees with the experimental data except in the 
region around $500\textrm{ MeV}$.
 The reason of this discrepancy will be further discussed.

\section{Results for the $\phi\to\pi^0\eta\gamma$ decay}

\begin{figure}
  \includegraphics[width=7cm,angle=-90]{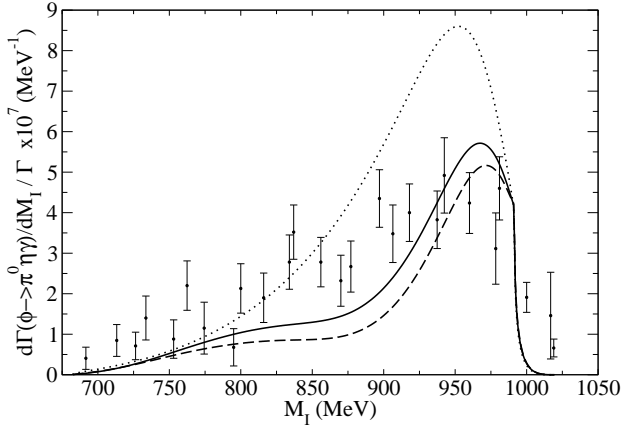}
\caption{\rm 
Different contributions to the $\pi^0\eta$ invariant mass
distributions of the $\phi\to\pi^0\eta\gamma$ decay:
Dotted line: chiral loops of Fig.~\ref{looplot}.
Dashed line: chiral loops of Fig.~\ref{looplot} + 
sequential VMD and its final
state interaction.
Solid line: idem plus the contribution of the mechanisms
involving axial-vector mesons, (full model).
}
\label{fig:res11}
\end{figure}

 After the discussion of the former points the consideration of the $\phi \rightarrow \pi^{0}\eta \gamma$ decay requires only minimal technical details which one can see in ref. \cite{ourphirad}.
We show in Fig. \ref{fig:res11} the results for the different
mechanisms in this reaction.

\begin{figure}
  \includegraphics[width=6cm,angle=-90]{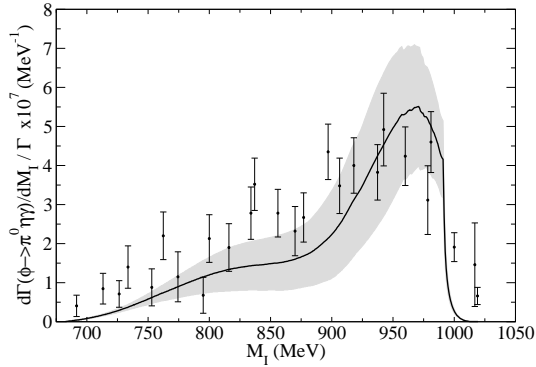}
\caption{\rm 
Final results for the $\pi^0\eta$ invariant mass distribution
for the $\phi\to\pi^0\eta\gamma$ decay with the theoretical
error band. }
\label{fig:res12}
\end{figure}

Again, in Fig.~\ref{fig:res12} we have plotted the
full model performing the
theoretical error analysis\footnote{
We have also checked that the use of a mixing angle for the
strange members of the axial nonets of around $30$ or $60$
degrees \cite{suzuki,axial} turns out in decreasing the lower
limit of the error band in around $5\%$ and $10\%$ for the 
$\phi\to\pi^0\pi^0\gamma$ and $\phi\to\pi^0\eta\gamma$ decays
respectively.
}.
We can see that when these   uncertainties are considered we
obtain a theoretical band
 in acceptable agreement with the
experimental data.

The total width and branching ratio obtained are

\begin{eqnarray}
BR(\phi\to\pi^0\eta\gamma)&=&(0.59\pm 0.19)\times 10^{-4}
\end{eqnarray}
to be compared with the experimental values
$BR^{exp}(\phi\to\pi^0\eta\gamma)
=(0.88\pm 0.14\pm 0.09)\times 10^{-4} 
$ \cite{novo3}, 
$(0.90\pm 0.24\pm 0.10)\times 10^{-4} 
$ \cite{novo2}, $
 (0.85\pm 0.05\pm 0.06)\times 10^{-4}
$ \cite{frascati2}.

\section{
Further discussion of the $\phi\to\pi^0\pi^0\gamma$ results.
}

We would like to comment on the strength that we obtain
around $500\textrm{ MeV}$ in the $\pi^0\pi^0$
 invariant mass distribution
in the $\phi\to\pi^0\pi^0\gamma$ decay which appears in contradiction with the experimental
analysis. As we saw it comes from accumulation of the novel
mechanisms which we have discussed in our paper. Such
mechanisms are not considered in other theoretical papers
which find a good agreement with the data, after fitting some parameters to the data. Our philosophy has been
different and we have not fitted any parameter to the $\phi$
radiative decay data but simple have considered the different
mechanisms that can sizeable contribute to the process. An
acceptable agreement with the data is found in the region of
the $f_0(980)$, which is the most important issue concerning
this reaction. This is not trivial a priori in view of the
very small width of the dynamically generated $f_0(980)$
(around $30\textrm{ MeV}$) that one obtains in the model of
\cite{npa} that we use here and the large 'visual' width of
the $f_0(980)$ peak in the present experiment. Part of the
reason for the agreement comes from the $q$ factor (photon
momentum) in the amplitude, as requirement by gauge
invariance and emphasize in \cite{achasovq,pennington}, which
gives more weight to the amplitude as we move down in the
$\pi\pi$ invariant mass from the upper limit (where $q=0$).
However, as seen in our results, the inclusion of the new
mechanisms and their interference with the dominant one,
particularly the contribution of the axial-vector meson
exchange mechanisms, also contributes to the widening of the
distribution around the $f_0(980)$ peak.

Although the agreement with  data at low masses is not very
good, we must point out two sources of uncertainty in the
experimental spectrum. First,  the results in the low and
intermediate mass region largely depend on  the background
subtraction dominated by the non-resonant $\omega\pi^0$ 
process. The size of this process is difficult to obtain
because  it has a strong background itself, mostly from the
$\phi\to f_0 \gamma$ process, as it is discussed in
\cite{kloe1}. There, its magnitude  has been obtained  in a
model dependent way assuming some a priori spectrum for the 
$\phi\to f_0 \gamma$ process \cite{kloe1}. In fact, before the
subtraction,  the raw data resemble much more our calculated
spectrum, (see fig. 4 from \cite{frascati1}), and we could
think of a slightly smaller $\omega\pi^0$ background.

Additionally, there is some  uncertainty in the way the data
are corrected to account for the experimental efficiency.
This is done in \cite{frascati1} by dividing the observed spectrum
by the effect of applying  the experimental efficiency on
some theoretical distribution. This unfolding
procedure depends on the theoretical model used, which  we 
think at low $\pi^0\pi^0$ masses is at least incomplete. 
 In fact, with the unfolding method used, the zero value 
of the spectrum obtained with the theoretical model of 
\cite{frascati1} implies unavoidably a zero value for the
corrected experimental results. A reanalysis to the light of the 
present discussion would be most welcome.


\section{$\sigma$ meson in a
nuclear medium through two pion photoproduction}

In the last years there has been an intense theoretical and
experimental debate about the nature of the $\sigma$ meson,
mostly centered on the discussion about its interpretation as an ordinary 
$q \bar{q}$ meson or a $\pi \pi $ resonance. 
The advent of $\chi PT$ showed up that the $\pi \pi$ interaction in s-wave
in the isoscalar sector is strong enough
to generate a resonance through multiple 
scattering of the pions. This seems to be the case, and even in models starting
with a seed of $q \bar{q}$ states, the incorporation of the $\pi \pi$
channels in a unitary approach leads to a large dressing by a pion
cloud which makes negligible the effects of the original $q \bar{q}$ seed.
This idea has been made more quantitative through the introduction
of the unitary extensions of $\chi PT$ ($U \chi PT$).
Even more challenging is the modification of the properties of the $\sigma$ meson at
finite nuclear density. Since present theoretical calculations agree on a sizeable modification in the
nuclear medium of the $\pi\pi$ scattering in the $\sigma$ region, our purpose here 
is to find out its possible experimental signature in a very suited process
 like the $(\gamma,\pi^0 \pi^0)$ reaction in nuclei. (This contribution is a summary of
 the more extended work \cite{Roca:2002vd}).
This reaction is much better suited  than the
$(\pi,\pi\pi)$ one to investigate the
modification of the $\pi\pi$ in nuclear matter
because the photons are not distorted
by the nucleus and the reaction can test higher densities.

\subsection{Model}
For the model of the elementary $(\gamma, \pi \pi)$ reaction we follow 
 \cite{Nacher:2000eq} which considers the coupling of the photons to mesons, 
 nucleons, and the resonances
$\Delta(1232)$, $N^*(1440)$, $N^*(1520)$ and $\Delta(1700)$.
This model relies upon tree level
diagrams. Final state interaction of the $\pi N$ system is accounted for
by means of the explicit use of resonances with their widths. However,
since we do not include explicitly the $\sigma$ resonance, the final state
interaction of the two pions has to be implemented to generate it.

The $\gamma N \to N \pi^0 \pi^0$ amplitude can be decomposed in
a part which has in the final state the combination of pions
in isospin I=0 and another part where the pions are in I=2.

\begin{eqnarray}
|\pi^0_1\pi^0_2>\,=\quad
\underbrace{\frac{1}{3}|\pi^0_1\pi^0_2+\pi^+_1\pi^-_2+\pi^-_1\pi^+_2>}
_{\textrm{I=0 part}}\quad  
\underbrace{-\frac{1}{3}|\pi^0_1\pi^0_2+\pi^+_1\pi^-_2+\pi^-_1\pi^+_2>
+|\pi^0_1\pi^0_2>}_{\textrm{I=2 part}}
\label{eq:T00}
\end{eqnarray}
  
The renormalization of the 
$I=0$ $(\gamma,\pi \pi)$ amplitude is done by factorizing the on shell 
tree level $\gamma N \to \pi \pi N $ and $\pi \pi \to \pi \pi$ amplitudes in the 
loop functions. 

\begin{equation}
T_{(\gamma,\pi^0\pi^0)}^{I_{\pi\pi}=0}\to T_{(\gamma,\pi^0\pi^0)}^{I_{\pi\pi}=0}
\left(1+G_{\pi\pi}t_{\pi\pi}^{I=0}(M_I)\right)
\label{eq:GT1}
\end{equation}
where $G_{\pi\pi}$ is the loop function of the two pion propagators, 
which appears in the Bethe Salpeter equation, and $t_{\pi\pi}^{I=0}$ is the
$\pi\pi$ scattering matrix in isospin I=0, taken from \cite{Chiang:1998di}.

\begin{figure}
  \includegraphics[height=.1\textheight]{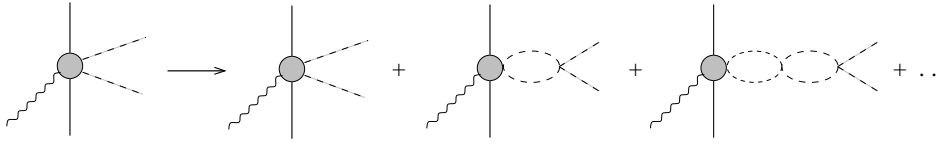}
\caption{\small{Diagrammatic series for pion final state interaction in I=0}}
\label{fig:Tgammapipi}
\end{figure}

In Fig.~\ref{fig:Tgammapipi} we show a diagrammatical
representation of the
the two pion production  including their final state
interaction.

The multiple scattering of the two final pions can be accounted for
by means of the Bethe Salpeter equation, 

\begin{equation}
t=V+VG_{\pi\pi}t
\label{eq:Bethe}
\end{equation}
where V is given by the lowest order chiral
amplitude for $\pi\pi\to\pi\pi$ in $I=0$
and $G_{\pi\pi}$, the loop function of the two pion propagators can be
regularized by means of a cut off or with
dimensional regularization. In both approaches it has been shown
that $V$ factorizes with its on shell value in the Bethe-Salpeter equation.
 Hence, in the Bethe-Salpeter equation the integral
involving $Vt$ and the product of the two pion propagators affects only these
latter two, since $V$ and $t$ factorize outside the integral, thus leading to
Eq.~(\ref{eq:Bethe}) where $VG_{\pi\pi}t$ is the algebraic product of V, the loop 
function of the two propagators, $G_{\pi\pi}$, and the $t$ matrix.

When we renormalize the I=0 amplitude in nuclei 
to account for the pion final state
interaction, we change $G$ and $t_{\pi\pi}^{I=0}$ by their corresponding results
in nuclear matter \cite{Chiang:1998di} evaluated at the local density.
In the model of \cite{Chiang:1998di},
the $\pi\pi$ rescattering in nuclear matter was
done renormalizing the pion propagators in the medium
and introducing vertex corrections for consistency.

In the model for $( \gamma,2\pi )$ of 
\cite{Nacher:2000eq} there are indeed contact terms as implied before, as well as
other terms involving intermediate nucleon states or resonances. In this
latter case the loop function involves three propagators but 
the intermediate baryon is far off shell an the factorization of
Eq.~(\ref{eq:GT1}) still holds. There is, however, an exception in
the $\Delta$ Kroll Ruderman term, since as we increase the photon energy we get
closer to the $\Delta$ pole. For this reason this term has
been dealt separately making the
explicit calculation of the loop with one $\Delta$ and two pion propagators.

The cross section for the process in nuclei is calculated using many body techniques.
From the imaginary part of the photon selfenergy diagram with a particle-hole
excitation and two pion lines
as intermediate states, the cross section can be expressed as

\begin{eqnarray}
\sigma=&&\frac{\pi}{k}\int d^3\vec{r}\int\frac{d^{3}\vec{p}}{(2\pi)^3}
\int\frac{d^{3}\vec{q_1}}{(2\pi)^3}
\int\frac{d^{3}\vec{q_2}}{(2\pi)^3}
\, F_1(\vec{r},\vec{q_1}) F_2(\vec{r},\vec{q_2})
\frac{1}{2\omega(\vec{q_1})}\frac{1}{2\omega(\vec{q_2})}
\nonumber\\%
&&\cdot
\sum_{s_i,s_f}\overline{\sum_{pol}}\mid T\mid^{2}n(\vec{p})[1-n(\vec{k}+\vec{p}
 -\vec{q_1}-\vec{q_2})]
\nonumber\\
&&\cdot\delta(k^{0}+E(\vec{p})-\omega(\vec{q_1})-\omega(\vec{q_2})-E(\vec{k}
+\vec{p}-\vec{q_1}-\vec{q_2}))\nonumber
 \label{eq:sigma2}
\end{eqnarray}
where the factors $F_i(\vec{r},\vec{q_i})$  take into account the distortion of
the final pions in their way out through the nucleus and are given by

 \begin{eqnarray}\hspace{0cm}
F_i(\vec{r},\vec{q_i})=exp\left[\int_{\vec{r}}^{\infty}dl_i \frac{1}{q_i}Im
\Pi (\vec{r}_i) \right]
\label{eq:Feikonal}
\\ \nonumber & &\vspace{0.4cm}\hspace{-5.5cm}
\vec{r_i}=\vec{r}+l_i \ \vec{q_i}/\mid \vec{q_i}\mid
\end{eqnarray}

where $\Pi$ is the pion selfenergy, taken from a model based on an extrapolation for low
energy pions of a pion-nucleus optical potential developed for pionic atoms using many
body techniques. The imaginary part of the potential is split into a part that
accounts for the probability of quasielastic collisions and another one which
accounts for the pion absorption.  With this approximation
the pions which undergo absorption are removed
from the flux but we do not remove those which undergo 
quasielastic collisions since they do not change in average
the shape or the strength of the $\pi\pi$ invariant mass distribution.

\subsection{Results}

In the figure we can see the results for the two pion
invariant mass distributions  in the $(\gamma,\pi^0\pi^0)$
 reaction on $^1H$, $^{12}C$
and $^{208}Pb$.  The difference between the  solid and dashed
curves is the use of the in medium $\pi \pi$ scattering and
$G$ function instead of the free ones, which we take from
\cite{Chiang:1998di}.  As one can see in the figure, there 
is an appreciable shift of strength to the low invariant mass
region due to the in medium  $\pi \pi$  interaction.   This
shift is remarkably similar to the one found in the
preliminary measurements of \cite{Messchendorp:2002au}.

These results show a clear signature of the modified $\pi\pi$
interaction in the medium.  The fact that the photons are not
distorted has certainly an advantage over the pion induced
reactions and allows one to see  inner parts of the nucleus.

Although we have been discussing the $\pi \pi$ interaction in
the nuclear medium it is clear that we can relate it to the
modification of the $\sigma$ in the medium. We have mentioned
that the reason for the shift of strength to lower invariant
masses in the mass distribution is due to the accumulated
strength in the scalar isocalar $\pi \pi$ amplitude in the
medium. Yet, this strength is mostly governed by the presence
of the $\sigma$ pole and there have been works suggesting
that the sigma should move to smaller masses and widths when
embedded in the nucleus. The
present results represent an evidence that the pole position of the
 $\sigma$  to smaller energies as the
nuclear density increases, a phenomenon which would come to
strengthen once more the nature of the $\sigma$ meson as
dynamically generated by the multiple scattering of the pions
through the underlying chiral dynamics.

\begin{figure}
  \includegraphics[height=.5\textheight]{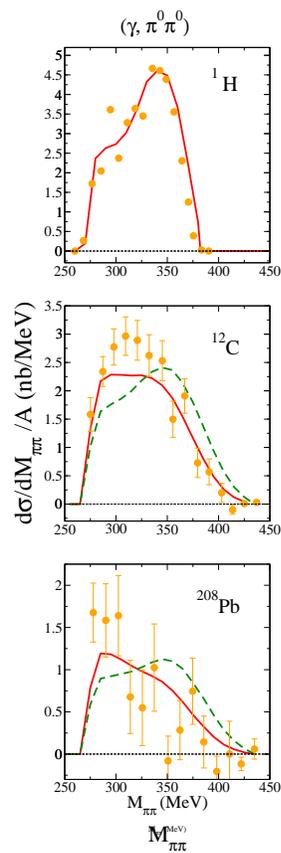}
\caption{\small{Two pion invariant mass distribution for $\pi^0\pi^0$
 photoproduction 
in $^{12}C$ and $^{208}Pb$.
Continuous lines: using the in medium final $\pi\pi$ interaction. 
Dashed lines: using the final $\pi\pi$ interaction at zero
density.
Exp. data from \cite{Messchendorp:2002au}}}
\end{figure}

\section{Conclusions}
We have demonstrated with two examples the relevance of chiral dynamics in
processes involving scalar mesons.  The main point to stress is that, since
the scalar mesons are generated dynamically in the scheme that we follow, we
do not face any unknown coupling and the theory is predictive for any process
involving the production of scalar mesons. The agreement with the data
obtained here and in many other processes \cite{report} gives strong support to the
picture of scalar mesons being dynamically generated by the multiple
scattering of the pseudoscalar mesons under the interaction provided by the
lowest order chiral Lagrangian.


\begin{thebibliography}{99}



\bibitem{greco} A. Bramon, G. Colangelo and M. Greco, Phys. Lett. B 287

\bibitem{franzini} P.J. Franzini, W. Kim and J. Lee Franzini, Phys. Lett.
 B 287 (1992) 259.

\bibitem{colangelo} G. Colangelo and P.J. Franzini, Phys. Lett. B 289
 (1992) 189.
 (1992) 263.

\bibitem{achasov} N.N. Achasov, V.V. Gabin and E.P. Solodov,
 Phys. Rev. D 55 (1997) 2672.

\bibitem{bramon5} A. Bramon, A. Grau and G. Pancheri, Phys. Lett. B 289 (1992) 97.

\bibitem{lucio}
A.~Bramon, R.~Escribano, J.~L.~Lucio M, M.~Napsuciale and G.~Pancheri,
Eur.\ Phys.\ J.\ C {\bf 26} (2002) 253
[arXiv:hep-ph/0204339].

\bibitem{uge}
E.~Marco, S.~Hirenzaki, E.~Oset and H.~Toki,
Phys.\ Lett.\ B {\bf 470} (1999) 20
[arXiv:hep-ph/9903217].

\bibitem{kyoto} Proc. of the Workshop on the " Posible existence of the
$\sigma$
meson and its implication in hadron physics", Kyoto June, 2000, Ed. S.
Ishida et
al., web page http://amaterasu.kek.jp/YITPws/.


\bibitem{npa}
J.~A.~Oller and E.~Oset,
Nucl.\ Phys.\ A {\bf 620} (1997) 438
[Erratum-ibid.\ A {\bf 652} (1999) 407]
[arXiv:hep-ph/9702314].

\bibitem{iam}
 J.A. Oller, E. Oset and J. R. Pel\'aez, Phys. Rev. Lett. 80 (1998) 3452;
 ibid, Phys. Rev. D 59 (1999) 74001.

\bibitem{nsd} J.A. Oller and E. Oset, Phys. Rev. D60 (1999) 074023

\bibitem{pertiou} J. Lucio and J. Pertiou, Phys. Rev. D 42 (1990) 3253;
 ibid D 43 (1991) 2447.

\bibitem{close}
F.~E.~Close, N.~Isgur and S.~Kumano,
Nucl.\ Phys.\ B {\bf 389} (1993) 513
[arXiv:hep-ph/9301253].

\bibitem{bramonnew}
A.~Bramon, R.~Escribano, J.~L.~Lucio M, M.~Napsuciale and G.~Pancheri,
Eur.\ Phys.\ J.\ C {\bf 26} (2002) 253
[arXiv:hep-ph/0204339].

\bibitem{novo1}
M.~N.~Achasov {\it et al.},
Phys.\ Lett.\ B {\bf 485} (2000) 349
[arXiv:hep-ex/0005017].
 
\bibitem{novo2}
R.~R.~Akhmetshin {\it et al.}  [CMD-2 Collaboration],
Phys.\ Lett.\ B {\bf 462} (1999) 380
[arXiv:hep-ex/9907006].

\bibitem{novo3}
M.~N.~Achasov {\it et al.},
Phys.\ Lett.\ B {\bf 479} (2000) 53
[arXiv:hep-ex/0003031].

\bibitem{frascati1} 
A.~Aloisio {\it et al.}  [KLOE Collaboration],
Phys.\ Lett.\ B {\bf 537} (2002) 21
[arXiv:hep-ex/0204013].

\bibitem{frascati2}
A.~Aloisio {\it et al.}  [KLOE Collaboration],
Phys.\ Lett.\ B {\bf 536} (2002) 209
[arXiv:hep-ex/0204012].


\bibitem{grau}
A.~Bramon, A.~Grau and G.~Pancheri,
Phys.\ Lett.\ B {\bf 283}, 416 (1992).

\bibitem{escribano}
A.~Bramon, R.~Escribano, J.~L.~Lucio Martinez and M.~Napsuciale,
Phys.\ Lett.\ B {\bf 517} (2001) 345
[arXiv:hep-ph/0105179].

\bibitem{palomar}
J.~E.~Palomar, S.~Hirenzaki and E.~Oset,
Nucl.\ Phys.\ A {\bf 707} (2002) 161
[arXiv:hep-ph/0111308].





\bibitem{achasovq}
N.~N.~Achasov and V.~N.~Ivanchenko,
Nucl.\ Phys.\ B {\bf 315} (1989) 465.

\bibitem{pennington}
M.~Boglione and M.~R.~Pennington,
arXiv:hep-ph/0303200.

\bibitem{neufeld}
K.~Huber and H.~Neufeld,
Phys.\ Lett.\ B {\bf 357} (1995) 221
[arXiv:hep-ph/9506257].

\bibitem{ecker}
G.~Ecker, J.~Gasser, A.~Pich and E.~de Rafael,
Nucl.\ Phys.\ B {\bf 321} (1989) 311.

\bibitem{oller}
J.~A.~Oller,
Phys.\ Lett.\ B {\bf 426} (1998) 7
[arXiv:hep-ph/9803214].

\bibitem{ourphirad}
J. E. Palomar, L. Roca, E. Oset and M. J. Vicente Vacas, in preparation.

\bibitem{achasovlast}
N.~N.~Achasov and A.~V.~Kiselev,
arXiv:hep-ph/0212153.



\bibitem{urech}
R.~Urech,
Phys.\ Lett.\ B {\bf 355} (1995) 308
[arXiv:hep-ph/9504238].

\bibitem{kozhe} N.N. Achasov {\it et al.}, Int. J. Mod. Phys. A7
  (1992) 3187, N.N. Achasov and A.A. Kozhevnikov, Phys. Lett.
  B233, (1989) 474 and Int. J. Mod. Phys. A7, (1992) 4825.

\bibitem{bramonepsi} 
A.~Bramon, R.~Escribano and M.~D.~Scadron,
Eur.\ Phys.\ J.\ C {\bf 7} (1999) 271
[arXiv:hep-ph/9711229].

\bibitem{bramonotro}
A.~Bramon, A.~Grau and G.~Pancheri,
Phys.\ Lett.\ B {\bf 283} (1992) 416.

\bibitem{bramonprivate} A.~Bramon, private communication 

\bibitem{roca}
E.~Oset, J.~R.~Pelaez and L.~Roca,
Phys.\ Rev.\ D {\bf 67} (2003) 073013
[arXiv:hep-ph/0210282].

\bibitem{pdg}
K.~Hagiwara {\it et al.}  [Particle Data Group Collaboration],
Phys.\ Rev.\ D {\bf 66} (2002) 010001.



\bibitem{gatto}
R.~Barbieri, R.~Gatto and Z.~Kunszt,
Phys.\ Lett.\ B {\bf 66} (1977) 349.

\bibitem{suzuki}
M.~Suzuki,
Phys.\ Rev.\ D {\bf 47} (1993) 1252.

\bibitem{carnegie}
R.~K.~Carnegie, R.~J.~Cashmore, W.~M.~Dunwoodie, T.~A.~Lasinski and D.~W.~Leith,
Phys.\ Lett.\ B {\bf 68} (1977) 287.

\bibitem{axial} L.~Roca, J.~E.~Palomar and E.~Oset. In
preparation.


\bibitem{kloe1}
S. Giovanella ans S. Miscetti, KLOE note 178



\bibitem{Roca:2002vd}
L.~Roca, E.~Oset and M.~J.~Vicente Vacas,
Phys.\ Lett.\ B {\bf 541} (2002) 77
[arXiv:nucl-th/0201054].

\bibitem{Nacher:2000eq}
J.~C.~Nacher, E.~Oset, M.~J.~Vicente and L.~Roca,
Nucl.\ Phys.\ A {\bf 695} (2001) 295
[arXiv:nucl-th/0012065].

\bibitem{Chiang:1998di}
H.~C.~Chiang, E.~Oset and M.~J.~Vicente-Vacas,
Nucl.\ Phys.\ A {\bf 644} (1998) 77
[arXiv:nucl-th/9712047].

\bibitem{Messchendorp:2002au}
J.~G.~Messchendorp {\it et al.},
Phys.\ Rev.\ Lett.\  {\bf 89} (2002) 222302
[arXiv:nucl-ex/0205009].

\bibitem{report}
J.~A.~Oller, E.~Oset and A.~Ramos,
Prog.\ Part.\ Nucl.\ Phys.\  {\bf 45} (2000) 157
[arXiv:hep-ph/0002193].

\end{thebibliography}
\end{document}